\documentclass[conference]{IEEEtran}
\IEEEoverridecommandlockouts

\usepackage{amsmath,amssymb,amsfonts}
\usepackage{algorithm}
\usepackage[noend]{algpseudocode}

\usepackage[font=small]{caption}
\usepackage{subcaption}
\usepackage{soul}
\usepackage{tikz}
\usepackage{enumitem}
\usepackage{calc}
\usepackage{flushend}
\usepackage{pgfplots}
\usepackage{booktabs}
\usepackage[a4paper, total={184mm,239mm}]{geometry}

\usepackage{pifont}%
\newcommand{\cmark}{\ding{51}}%
\newcommand{\xmark}{\ding{55}}%

\definecolor{codegreen}{rgb}{0,0.6,0}
\definecolor{forestgreen(web)}{rgb}{0.13, 0.55, 0.13}

\definecolor{jcgreen}{HTML}{4daf4a}
\definecolor{jcblue}{HTML}{377eb8}
\definecolor{jcred}{HTML}{e41a1c}

\begin{document}
\title{PASS: Exploiting Post-Activation Sparsity in Streaming Architectures for CNN Acceleration}
\author{
     \IEEEauthorblockN{Alexander Montgomerie-Corcoran\IEEEauthorrefmark{1}, Zhewen Yu\IEEEauthorrefmark{1}, Jianyi Cheng and Christos-Savvas Bouganis}
    \IEEEcompsocitemizethanks{\IEEEcompsocthanksitem\IEEEauthorrefmark{1}\textit{equal contribution}}
     \IEEEauthorblockA{
     Imperial College London, UK \\
     \{alexander.montgomerie-corcoran15, zhewen.yu18, jianyi.cheng17, christos-savvas.bouganis\}@imperial.ac.uk}
}

\maketitle
\begin{abstract}
With the ever-growing popularity of Artificial Intelligence, there is an increasing demand for more performant and efficient underlying hardware.
Convolutional Neural Networks (CNN) are a workload of particular importance, which achieve high accuracy in computer vision applications. Inside CNNs, a significant number of the post-activation values are zero, resulting in many redundant computations.
Recent works have explored this post-activation sparsity on instruction-based CNN accelerators but not on streaming CNN accelerators, despite the fact that streaming architectures are considered the leading design methodology in terms of performance.
In this paper, we highlight the challenges associated with exploiting post-activation sparsity for performance gains in streaming CNN accelerators, and demonstrate our approach to address them. 
Using a set of modern CNN benchmarks, our streaming sparse accelerators achieve 1.41$\times$ to 1.93$\times$ efficiency (GOP/s/DSP) compared to state-of-the-art instruction-based sparse accelerators.

\end{abstract}

\section{Introduction}
\label{sec:introduction} 
Despite the successes of Transformers \cite{vit}, CNNs \cite{howard2019searching, ding2021repvgg} are still the de-facto method for many vision tasks. 
The Rectified Linear Unit (ReLU) activation operation is ubiquitous within CNNs as it introduces non-linearities, increasing the network capacity.
ReLU operation clamps all negative values in the feature maps to zero, leading to many redundant operations, a property referred to as post-activation sparsity.
As an alternative to post-activation sparsity, weight sparsity can be induced through the pruning of near-zero parameters \cite{li2019squeezeflow}.
However, weight sparsity exploitation incurs an accuracy penalty, and requires re-training to recover accuracy losses. 
On the contrary, post-activation sparsity exploitation does not introduce approximations in the computations, and offers a large potential for performance gains.

Recent works have explored post-activation sparsity on instruction-based CNN accelerators \cite{gondimalla2019sparten, zhu_efficient_2020}, where the computation of layers is time-multiplexed in hardware.
As feature maps are repeatedly read to and from off-chip memory during execution, they are encoded to reduce memory footprint.
In terms of dense CNN accelerator designs, streaming architectures are considered more efficient \cite{venieris_toolflows_2018}, as the layerwise hardware customisation and pipelining leads to a lower memory footprint and higher throughput. 
However, of all the prior streaming architecture works none have exploited post-activation sparsity.
Streaming architectures pose the following challenges with regard to post-activation sparsity:
\begin{itemize}
    \item \textit{Dynamic Scheduling}: 
        As the appearance of zeros is only known at run-time, a dynamic scheduler \cite{cheng2022dass,josipovic2022dataflow} is required to skip zeros and dispatch the remaining non-zero elements to computation units on-the-fly.
    \item \textit{Data Stream Synchronisation}: 
        Multiple data streams are processed in parallel for high throughput. 
        However, the density and distribution of zeros across the feature maps vary, leading to an imbalanced workload between data streams which causes pipeline stalls. Load balancing and buffering are required.
    \item \textit{Resource Allocation}: 
        The allocation of FPGA resources to the various compute engines for a given performance target is a non-convex problem. Existing algorithms \cite{montgomerie2022samo} provide solutions for the case of multiple dense computation engines, but sparse computation engines have not yet been considered. 
\end{itemize}

\begin{figure}[t]
    \centering
    \hspace{-2.5cm}
    \resizebox{1.1\columnwidth}{!}{\definecolor{sparsecolor}{HTML}{b2df8a}
\definecolor{densecolor}{HTML}{1f78b4}
\definecolor{pgfred}{HTML}{e31a1c}

\definecolor{myblue}{rgb}{0.35, 0.45, 0.64}
\definecolor{myorange}{rgb}{0.79, 0.53, 0.38}
\definecolor{myred}{rgb}{0.7, 0.36, 0.37}
\definecolor{mygreen}{rgb}{0.37, 0.61, 0.42}
\definecolor{mymagenta}{rgb}{0.52, 0.47, 0.66}
\definecolor{myyellow}{rgb}{1.0, 0.75, 0.0}
\definecolor{mycyan}{rgb}{0.0, 0.72, 0.92}

\usetikzlibrary{patterns}

\begin{tikzpicture}
\pgfplotsset{compat=1.5}
\begin{axis}[
    ybar,
    height=55mm,
    width=\columnwidth,
    bar width=7,
    ticklabel style = {font=\footnotesize},
    ylabel style={align=center}, 
    ylabel={\small Performance Density},
    xtick={1,2,3,4,5,6,7,8,9,10,11},
    x tick style={draw=none},
    xticklabels={
        \cite{venieris_fpgaconvnet_2018}, \cite{li2021fpga}, \cite{lu2019efficient}, \cite{zhu_efficient_2020}, Ours, Ours, \cite{venieris_fpgaconvnet_2018}, \cite{venieris2021unzipfpga}, Ours, \cite{lu2019efficient}, Ours
    },
    ymin=0.15, ymax=0.75,
    xmin=0.5, xmax=11.5,
    every axis plot/.append style={
        ybar,
        bar shift=0pt,
        fill
    },
    legend style={at={(0.5,1.15)}, anchor=north,legend columns=-1, font=\scriptsize},
    legend image code/.code={%
        \draw[#1] (0cm,-0.05cm) rectangle (0.4cm,0.1cm);
    },
    legend entries={ZC706, VC709, ZCU102},
  ]

\node[] at (30, 550) {\footnotesize \textbf{VGG16}};
\node[] at (75, 550) {\footnotesize \textbf{ResNet-18}};
\node[] at (100, 550) {\footnotesize \textbf{ResNet-50}};

\addlegendimage{fill=white}
\addlegendimage{fill=white, pattern=crosshatch}
\addlegendimage{fill=white, pattern=north west lines}

\addplot[fill=jcblue!75 , draw=none, ] coordinates {(1, 0.33)};
\addplot[fill=jcgreen, draw=none, preaction={fill, jcblue!75}, pattern=crosshatch] coordinates {(2, 0.35)};
\addplot[fill=jcblue , draw=none, preaction={fill, jcblue!75}, pattern=north west lines] coordinates {(3, 0.27)};
\addplot[fill=jcblue , draw=none, preaction={fill, jcblue!75}, pattern=north west lines] coordinates {(4, 0.37)};
\addplot[fill=jcred!75  , draw=none, ] coordinates {(5, 0.61)};
\addplot[fill=jcblue , draw=none, preaction={fill, jcred!75}, pattern=north west lines ] coordinates {(6, 0.52)};
\addplot[fill=jcblue!75 , draw=none, ] coordinates {(7, 0.23)};
\addplot[fill=jcblue!75 , draw=none, ] coordinates {(8, 0.20)};
\addplot[fill=jcred!75  , draw=none, ] coordinates {(9, 0.35)};
\addplot[fill=jcblue , draw=none, preaction={fill, jcblue!75}, pattern=north west lines ] coordinates {(10, 0.25)};
\addplot[fill=jcred  , draw=none, preaction={fill, jcred!75}, pattern=north west lines  ] coordinates {(11, 0.24)};

\draw [dashed] (axis cs:6.5,0) -- (axis cs:6.5,1.0);
\draw [dashed] (axis cs:9.5,0) -- (axis cs:9.5,1.0);

\edef\mylst{ "1.22" , "1.30" , "1.0" , "1.37" , "2.59" , "1.93" , "1.15" , "1.0" , "1.75" , "1.0" , "0.96" }

\node[above,anchor=west, rotate=90] at (axis cs:1, 0.33+0.005) {\scriptsize $1.22\times$};
\node[above,anchor=west, rotate=90] at (axis cs:2, 0.35+0.005) {\scriptsize $1.30\times$};
\node[above,anchor=west, rotate=90] at (axis cs:3, 0.27+0.005) {\scriptsize $1.0\times$};
\node[above,anchor=west, rotate=90] at (axis cs:4, 0.37+0.005) {\scriptsize $1.37\times$};
\node[above,anchor=west, rotate=90] at (axis cs:5, 0.61+0.005) {\scriptsize $2.59\times$};
\node[above,anchor=west, rotate=90] at (axis cs:6, 0.52+0.005) {\scriptsize $1.93\times$};
\node[above,anchor=west, rotate=90] at (axis cs:7, 0.23+0.005) {\scriptsize $1.15\times$};
\node[above,anchor=west, rotate=90] at (axis cs:8, 0.20+0.005) {\scriptsize $1.0\times$};
\node[above,anchor=west, rotate=90] at (axis cs:9, 0.35+0.005) {\scriptsize $1.75\times$};
\node[above,anchor=west, rotate=90] at (axis cs:10, 0.25+0.005) {\scriptsize $1.0\times$};
\node[above,anchor=west, rotate=90] at (axis cs:11, 0.24+0.005) {\scriptsize $0.96\times$};

\end{axis}
\end{tikzpicture}}
    \caption{
            Comparison of our work against existing dense and sparse accelerators in terms of  performance density (GOP/s/DSP).
            Details can be found in Table~\ref{tab:main-comparison}.
        }
    \label{fig:perf-density}
    \vspace{-0.4cm}
\end{figure}
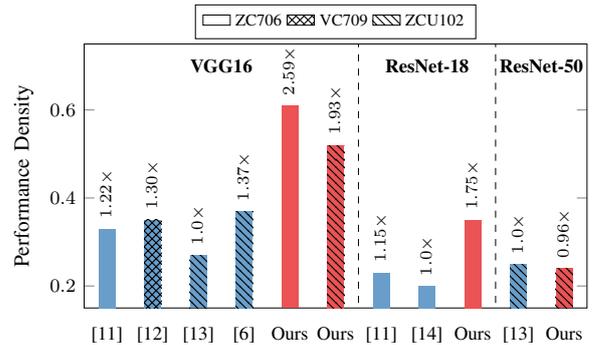

In this work, a toolflow that generates streaming architecture is proposed which addresses the preceding research questions.
The major contributions of our work include:
\begin{itemize}
    \item A scalable, dynamically scheduled architecture design which can exploit post-activation sparsity.
    \item A compile-time workload balancing strategy and automated buffer sizing methodology to reduce pipeline stalls.
    \item A novel Design Space Exploration (DSE) method which considers measured sparsity distribution for balancing the rates of multiple sparse engines. 
\end{itemize}
Over a set of CNN benchmarks, the results show that our approach can achieve 1.52$\times$ to 1.85$\times$ performance density compared to existing streaming dense architecture, and 1.41$\times$ to 1.93$\times$ performance density compared to state-of-the-art instruction-based sparse accelerators.

\begin{table}[t]
    \centering
    \caption{Comparison between our work and related works. Our work is the first to exploit post-activation sparsity on streaming architectures, with a novel DSE method that considers sparsity statistics. We also avoid the overhead of sparse encoding with dynamic scheduling.}
    \label{tab:high-level-comparison}
    \begin{tabular}{ccccccccc}
    \toprule 
       Approach  & \cite{venieris2021unzipfpga} &\cite{sharma2016dnnweaver} & \cite{sohrabizadeh2020end} & \cite{venieris_fpgaconvnet_2018} & \cite{zhu_efficient_2020} & \cite{li2021fpga} & \cite{lu2019efficient} & Ours \\
    \midrule
        Streaming & \color{jcred}{\xmark} & \color{jcred}{\xmark} & \color{jcgreen}{\cmark} & \color{jcgreen}{\cmark} & \color{jcred}{\xmark}   & \color{jcred}{\xmark} & \color{jcred}{\xmark} & \color{jcgreen}{\cmark} \\
        Sparsity  & \color{jcred}{\xmark} & \color{jcred}{\xmark} & \color{jcred}{\xmark} & \color{jcred}{\xmark} & \color{jcgreen}{\cmark} & \color{jcgreen}{\cmark} & \color{jcgreen}{\cmark}   & \color{jcgreen}{\cmark} \\
        Encoding  & \color{jcred}{\xmark} & \color{jcred}{\xmark} & \color{jcred}{\xmark} & \color{jcred}{\xmark} & \color{jcgreen}{\cmark} & \color{jcgreen}{\cmark} & \color{jcred}{\xmark}  & \color{jcred}{\xmark} \\
        DSE       & \color{jcgreen}{\cmark} & \color{jcgreen}{\cmark} & \color{jcred}{\xmark} & \color{jcgreen}{\cmark} & \color{jcred}{\xmark} & \color{jcred}{\xmark} & \color{jcred}{\xmark}  & \color{jcgreen}{\cmark} \\
    \bottomrule
    \end{tabular}
    \vspace{-0.3cm}
\end{table}

\section{Related Works}
\label{sec:related_work}

CNN Accelerators can generally be placed into two categories \cite{venieris_toolflows_2018}: instruction-based architectures and streaming architectures.
Instruction-based architectures \cite{jouppi2017datacenter, liu2021leveraging} use micro-instructions to execute layers of the CNN model, typically with a systolic array hardware design for computation.
These architectures are able to generalise across multiple CNN models, however, are limited by the inefficient processor-like control mechanisms.
Streaming architectures are characterised by their deeply pipelined hardware, where every layer of the CNN is mapped to a dedicated computation engine. 
FPGAConvNet \cite{venieris_fpgaconvnet_2018}, FINN \cite{blott2018finn} and HLS4ML \cite{duarte_fast_2018} are examples of toolflows which automate the generation of streaming architecture designs for a given target platform.
This design method often achieves high-performance designs, as the workload is tailored to the available resources of the platform.
However, these existing tools do not exploit post-activation sparsity as they are statically scheduled.

Sparse CNN accelerators have been explored in several works, especially on weight sparsity \cite{you_rsnn_2018, lu_spwa_2018}, as the deterministic patterns of non-zero operations can be scheduled in a static manner.
However, when dealing with post-activation sparsity, the presence of zeros is neither evenly distributed nor compile-time known, which poses a unique challenge for hardware designs. 
Existing research into exploiting post-activation sparsity has focused on two directions: skipping zero operations for energy savings, and using sparse representations for performance gains.
For energy saving, zero-gating unnecessary computation to leave the arithmetic circuitry idle leads to greater energy efficiency \cite{zhu_efficient_2020}, however, there are no performance benefits.
Works that do exploit post-activation sparsity for performance gains require the data to be encoded in a sparse representation \cite{scnn, sparten}.
By using representations such as Compressed Sparse Row (CSR) \cite{scnn}, non-zero operations can be routed to arithmetic circuits in a fine-grained manner.
However, the overheads associated with sparse encoding are not scalable, making this approach only suited to extremely sparse workloads \cite{liang2020omni}.
Furthermore, this approach has only been explored with instruction-based architectures, whose underlying hardware does not have the same performance benefits and challenges as streaming architectures.
This work is the first to address the challenge of exploiting post-activation sparsity for performance gains in streaming architectures whilst utilising the original dense representation of the feature maps by designing a novel matrix-vector hardware which is dynamically scheduled.

\section{Architecture}
\label{sec:architecture}

In this section, the architecture of the proposed accelerator is described.
The hardware design follows a streaming architecture approach.

\begin{figure}[t]
	\centering
	\includegraphics[width=0.65\columnwidth]{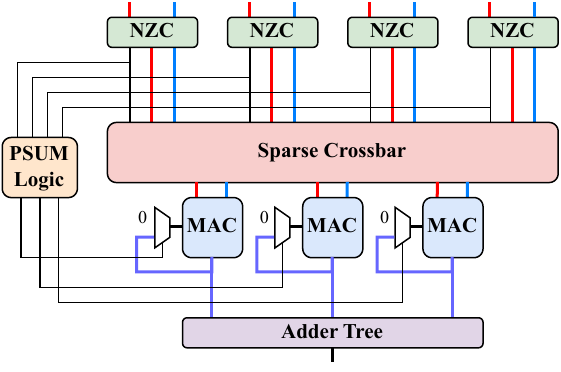}
	\caption{
            Diagram of the proposed Sparse Matrix-Vector Engine for $K_{\text{\tt x}}, K_{\text{\tt y}}=2$ and $k=3$.
            The \textit{Non-Zero Check (NZC)} hardware generates a signal indicating if the feature map is non-zero.
        }
	\label{fig:sparse_vector_dot_diagram}
\end{figure}

\subsection{Sparse Matrix-Vector Engine}
\label{sec:smve-hardware}

The key operation being performed in streaming architecture accelerators is matrix-vector multiplication.
In order to exploit sparsity in matrix-vector multiplication for performance gains, a novel Sparse Matrix-Vector Engine (S-MVE) is proposed.
A diagram of the proposed hardware is given in Fig.~\ref{fig:sparse_vector_dot_diagram}. 
The hardware accepts streams of consecutive pairs of vectors from the sliding window module and weights memory.
Each pair of feature map and weight elements is evaluated with a \textit{Non-Zero Check (NZC)} module, which indicates the presence of non-zero inputs.
The vector pairs are then fed into a \textit{Sparse Crossbar} module along with the \textit{NZC} flags.
This module squeezes the $K_{\text{\tt x}} \times K_{\text{\tt y}}$ inputs to $k$ outputs, and only routes the non-zero values to the crossbar output.
The crossbar output feeds the non-zero values to $k$ parallel MAC units, which perform the non-zero products.
The products are then accumulated through an adder tree.
Additional logic is required to handle extremely dense inputs, where the accumulation takes multiple cycles.

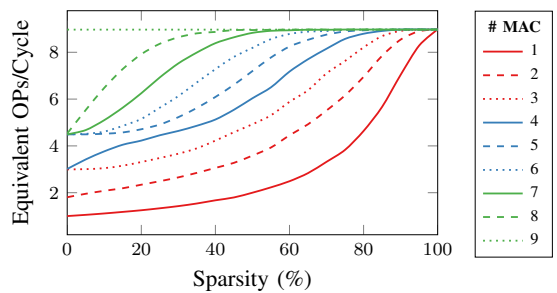
\begin{figure}[h]
	\centering
        \resizebox{0.85\columnwidth}{!}{\definecolor{sparsecolor}{HTML}{b2df8a}
\definecolor{densecolor}{HTML}{1f78b4}
\definecolor{pgfred}{HTML}{e31a1c}

\begin{tikzpicture}
\pgfplotsset{compat=1.5}
\begin{axis}[
  width=0.8\columnwidth,
  height=50mm,
  xlabel=Sparsity (\%),
  ylabel=Equivalent OPs/Cycle,
  tick label style={font=\footnotesize},  
  xmin=0, xmax=100,
  legend style={at={(1.1,1)},anchor=north west, font=\scriptsize},
]
\addlegendimage{empty legend}

\addplot [smooth, jcred, thick] 
coordinates {
    (00, 1.0017363429945239)
    (05, 1.0541110330288124)
    (10, 1.112924766285799)
    (15, 1.1799874134675896)
    (20, 1.2502083680613436)
    (25, 1.3345195729537367)
    (30, 1.4270311409906766)
    (35, 1.539198248734437)
    (40, 1.6752289479562208)
    (45, 1.7974117271129573)
    (50, 2.0003556187766716)
    (55, 2.2294887039239)
    (60, 2.4903154399557277)
    (65, 2.8369688563863322)
    (70, 3.3200531208499338)
    (75, 3.8310914353822576)
    (80, 4.638218923933209)
    (85, 5.668934240362812)
    (90, 7.0356472795497185)
    (95, 8.287292817679559)
    (100, 8.978451715881883)
};
\addplot [jcred, smooth, dashed, thick] 
coordinates {
    (00, 1.8069386443944748)
    (05, 1.954482279360667)
    (10, 2.0815986677768525)
    (15, 2.190847127555988)
    (20, 2.3422860712054967)
    (25, 2.4914184475694827)
    (30, 2.655180552277555)
    (35, 2.8484618306114697)
    (40, 3.060391730141458)
    (45, 3.275585965933906)
    (50, 3.585657370517928)
    (55, 3.930817610062893)
    (60, 4.452800316643578)
    (65, 4.923413566739606)
    (70, 5.513354569958343)
    (75, 6.1593211059403234)
    (80, 6.950880444856349)
    (85, 7.8506629448709)
    (90, 8.542141230068337)
    (95, 8.900316455696203)
    (100, 8.978451715881883)
};
\addplot [jcred, smooth, dotted, thick] 
coordinates {
    (00, 2.9972026108964966)
    (05, 3.006413682522715)
    (10, 3.048780487804878)
    (15, 3.1547952888390354)
    (20, 3.3117456579334705)
    (25, 3.483511379470506)
    (30, 3.6662864591820106)
    (35, 3.896103896103896)
    (40, 4.226145755071375)
    (45, 4.585286325657225)
    (50, 4.9244911359159556)
    (55, 5.307855626326964)
    (60, 5.882352941176471)
    (65, 6.377551020408164)
    (70, 7.100031555695803)
    (75, 7.62970498474059)
    (80, 8.211678832116789)
    (85, 8.690614136732329)
    (90, 8.914421553090333)
    (95, 8.967716221602233)
    (100, 8.974870362983646)
};
\addplot [jcblue, smooth, thick] 
coordinates {
    (00, 3.0132583366813983)
    (05, 3.4220532319391634)
    (10, 3.7663207231335787)
    (15, 4.05040504050405)
    (20, 4.226939695660342)
    (25, 4.454563452781628)
    (30, 4.639175257731959)
    (35, 4.860661049902787)
    (40, 5.126452494873548)
    (45, 5.5432372505543235)
    (50, 6.035407725321888)
    (55, 6.502890173410405)
    (60, 7.174744897959184)
    (65, 7.702841492639507)
    (70, 8.152173913043478)
    (75, 8.571428571428571)
    (80, 8.792497069167643)
    (85, 8.910891089108912)
    (90, 8.971291866028707)
    (95, 8.974870362983646)
    (100, 8.974870362983646)
};
\addplot [jcblue, smooth, dashed, thick] 
coordinates {
    (00, 4.492811501597444)
    (05, 4.493708807669263)
    (10, 4.519887505022097)
    (15, 4.575030500203335)
    (20, 4.711055276381909)
    (25, 4.913736623716969)
    (30, 5.227695167286245)
    (35, 5.622188905547226)
    (40, 6.090958310774228)
    (45, 6.611813106082868)
    (50, 7.197696737044146)
    (55, 7.777393708952644)
    (60, 8.244778307072188)
    (65, 8.532423208191126)
    (70, 8.754863813229573)
    (75, 8.889766890557093)
    (80, 8.939213349225268)
    (85, 8.960573476702509)
    (90, 8.971291866028707)
    (95, 8.971291866028707)
    (100, 8.971291866028707)
};
\addplot [jcblue, smooth, dotted, thick] 
coordinates {
    (00, 4.492811501597444)
    (05, 4.514446227929374)
    (10, 4.610655737704918)
    (15, 4.8564644938484784)
    (20, 5.152278452026563)
    (25, 5.633450175262895)
    (30, 6.125782738905527)
    (35, 6.712410501193317)
    (40, 7.279197670656745)
    (45, 7.804370447450572)
    (50, 8.187772925764191)
    (55, 8.564902931100114)
    (60, 8.782201405152225)
    (65, 8.88274772996447)
    (70, 8.907363420427554)
    (75, 8.960573476702509)
    (80, 8.971291866028707)
    (85, 8.971291866028707)
    (90, 8.971291866028707)
    (95, 8.971291866028707)
    (100, 8.971291866028707)
};
\addplot [jcgreen, smooth, thick] 
coordinates {
    (00, 4.492811501597444)
    (05, 4.675810473815462)
    (10, 5.099728014505893)
    (15, 5.634861006761833)
    (20, 6.25)
    (25, 6.927339901477833)
    (30, 7.525083612040134)
    (35, 7.990056818181818)
    (40, 8.389261744966444)
    (45, 8.640552995391705)
    (50, 8.830455259026687)
    (55, 8.9179548156956)
    (60, 8.939213349225268)
    (65, 8.96414342629482)
    (70, 8.960573476702509)
    (75, 8.971291866028707)
    (80, 8.971291866028707)
    (85, 8.971291866028707)
    (90, 8.971291866028707)
    (95, 8.971291866028707)
    (100, 8.971291866028707)
};
\addplot [jcgreen, smooth, dashed, thick] 
coordinates {
    (00, 4.5308095046314945)
    (05, 5.533694048204624)
    (10, 6.4804147465437785)
    (15, 7.286269430051814)
    (20, 7.916959887403237)
    (25, 8.364312267657992)
    (30, 8.640552995391705)
    (35, 8.809710258418168)
    (40, 8.868742609381158)
    (45, 8.949880668257757)
    (50, 8.949880668257757)
    (55, 8.960573476702509)
    (60, 8.967716221602233)
    (65, 8.971291866028707)
    (70, 8.971291866028707)
    (75, 8.971291866028707)
    (80, 8.971291866028707)
    (85, 8.971291866028707)
    (90, 8.971291866028707)
    (95, 8.971291866028707)
    (100, 8.971291866028707)
};
\addplot [jcgreen, smooth, dotted, thick] 
coordinates {
    (00, 8.967716221602233)
    (05, 8.967716221602233)
    (10, 8.967716221602233)
    (15, 8.967716221602233)
    (20, 8.967716221602233)
    (25, 8.967716221602233)
    (30, 8.967716221602233)
    (35, 8.967716221602233)
    (40, 8.967716221602233)
    (45, 8.967716221602233)
    (50, 8.967716221602233)
    (55, 8.967716221602233)
    (60, 8.967716221602233)
    (65, 8.967716221602233)
    (70, 8.967716221602233)
    (75, 8.967716221602233)
    (80, 8.967716221602233)
    (85, 8.967716221602233)
    (90, 8.967716221602233)
    (95, 8.967716221602233)
    (100, 8.967716221602233)
};
\addlegendentry{\hspace{-.6cm}\textbf{\# MAC}}
\addlegendentry{1}
\addlegendentry{2}
\addlegendentry{3}
\addlegendentry{4}
\addlegendentry{5}
\addlegendentry{6}
\addlegendentry{7}
\addlegendentry{8}
\addlegendentry{9}
\end{axis}
\end{tikzpicture}}
	\caption{
            Performance of the proposed Sparse Matrix-Vector Engine against sparsity for a typical $K_{\text{\tt x}}, K_{\text{\tt y}}=3$ kernel size, across all MAC configurations. With the increasing sparsity, fewer MACs are required to achieve the maximum performance.
        }
	\label{fig:svme-perf}
\end{figure}

The performance characteristics of the proposed S-MVE are given in Fig.~\ref{fig:svme-perf} for a typical $K_{\text{\tt x}}, K_{\text{\tt y}}=3$ kernel size, across all the different MAC configurations.
It can be seen that allocating more MACs to our engine leads to greater performance, but with increasing sparsity, the performance gain per MAC diminishes. 
For a sparsity level greater than 40\%, our implementation is always more resource-efficient than an equivalent dense implementation, achieving the maximum performance with fewer MACs allocated.
Therefore, our hardware exposes a fine-grained trade-off between performance and MAC resources at different sparsity levels.

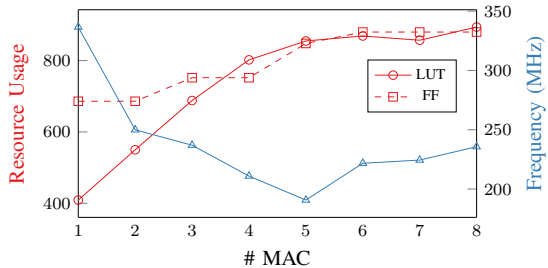
\begin{figure}[h]
\centering
    \hspace{-0.5cm}
    \resizebox{0.85\columnwidth}{!}{\definecolor{sparsecolor}{HTML}{b2df8a}
\definecolor{densecolor}{HTML}{1f78b4}
\definecolor{pgfjcred}{HTML}{e31a1c}

\begin{tikzpicture}
\pgfplotsset{width=0.9\columnwidth, height=50mm, compat=1.5}
\begin{axis}[
  axis y line*=left,
  xlabel=\# MAC,
  ylabel={\textcolor{jcred}{Resource Usage}},
  xtick style={draw=none},
  tick label style={font=\footnotesize},  
  xmin=1, xmax=8,
  legend style={at={(0.95,0.76)}, anchor=north east, font=\scriptsize},  
]
\addplot [jcred, mark=o] 
coordinates {
    (1, 409)
    (2, 550)
    (3, 688)
    (4, 802)
    (5, 855)
    (6, 869)
    (7, 857)
    (8, 894)
};
\addplot [jcred, mark=square, dashed, mark options={solid}] 
coordinates {
    (1, 686)
    (2, 686)
    (3, 752)
    (4, 752)
    (5, 848)
    (6, 880)
    (7, 880)
    (8, 880)
};
\legend{LUT, FF}
\end{axis}
\begin{axis}[
  ylabel={\textcolor{jcblue}{Frequency (MHz)}},
  axis y line*=right,
  axis x line=none,
  xmin=1, xmax=8, 
  tick label style={font=\footnotesize},  
]
\addplot [jcblue, mark=triangle] 
coordinates {
    (1, 336.58700774150117) 
    (2, 249.93751562109472)
    (3, 236.96682464454972) 
    (4, 210.92596498628984) 
    (5, 190.69412662090008)
    (6, 221.72949002217297)
    (7, 224.41651705565528)
    (8, 235.68230025925052)
};
\end{axis}
\end{tikzpicture}}
    \caption{
        LUT and FF resource usage as well as achieved clock frequency for a typical $K_{\text{\tt x}}, K_{\text{\tt y}}=3$ kernel size, across all MAC configurations, synthesised for a Zynq-Ultrascale+ FPGA architecture. 
        The design is able to maintain a frequency above 190MHz for all MAC configurations.
    }
    \label{fig:svme_rsc_freq}
\end{figure}

Furthermore, the utilisation characteristics and achieved clock frequency of the S-MVE hardware are explored in Fig.~\ref{fig:svme_rsc_freq}.
For the proposed S-MVE, all the configurations are able to achieve a clock frequency above 190MHz, with up to 340MHz for extremely sparse hardware.
The frequency dips towards the middle configuration, as this configuration contains the most complex crossbar with regard to routing.
There is a steady increase in LUT and FF resources as the number of allocated MACs increases, however this relationship plateaus around the 5-MAC configuration.
The LUT overheads mainly come from the sparse crossbar, and are not significant considering the savings on MACs. 
For reference, the cost of implementing a 16-bit MAC is 305 LUTs for the given FPGA fabric.

\subsection{Pipelined Convolutional Layer}

\begin{figure}[t]
    \centering
    \includegraphics[trim={0 0 0 0.5cm},clip, width=\columnwidth]{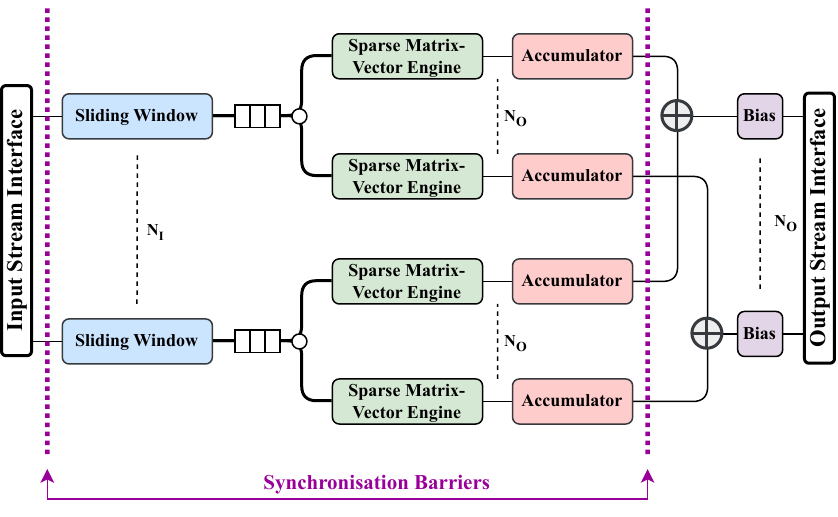}
    \caption{
        A block diagram of a convolutional layer within a streaming architecture design. 
        The input channel parallelism ($N_{\text{\tt I}}$) and output channel parallelism ($N_{\text{\tt O}}$) are highlighted. 
        The synchronisation boundaries required are also shown.
    }
    \label{fig:streaming-diagram}
\end{figure}

Fig.~\ref{fig:streaming-diagram} illustrates the pipelined design inside a single convolutional layer as well as the integration of the S-MVE module.  
The hardware components are as follows: 
\begin{itemize}
    \item \textit{Sliding Window}, which generates windows of the feature map from a single incoming stream by using line buffers.
    \item \textit{Sparse Matrix-Vector Engine} takes as input the incoming feature map windows and corresponding weights, and performs the non-zero dot-products within the kernel dimensions of the convolution.
    \item \textit{Accumulator}, which sums across the channel dimension.
    \item \textit{Bias Module}, which adds a per-channel bias term.
\end{itemize}

Apart from the kernel parallelism within the S-MVE module denoted by $k$, the convolutional layer hardware also exploits input-channel and output-channel parallelism with the factors denoted as $N_{\text{\tt I}}$ and $N_{\text{\tt O}}$ respectively in the Fig.~\ref{fig:streaming-diagram}. 
Overall, there are $N_{\text{\tt I}} \cdot N_{\text{\tt O}}$ S-MVE modules computing in parallel within each convolutional layer. 
The communication between S-MVE modules is asynchronous as the instantaneous sparsity can be different across data streams, therefore synchronisation barriers are required, as illustrated in the figure.

\section{Design Space Exploration}
\label{sec:method}
\begin{table}[t]
    \centering
    \caption{Taxonomy of symbols used in this paper.}
    \label{tab:symbols}
    \small
    \begin{tabular}{ll}
    \toprule
    \textbf{Symbols} & \textbf{Definitions} \\
    \midrule
        $B$ & batch size of feature map \\ 
        $L$ & total layers in the network \\ 
        $C_{\text{\tt I}}, C_{\text{\tt O}}$ & number of channels into and out of the layer \\ 
        $H_{\text{\tt O}}, W_{\text{\tt O}}$ & spatial dimensions of the output feature map \\ 
        $K_{\text{\tt x}}, K_{\text{\tt y}}$ & height and width of convolution kernel \\ 
        $N_{\text{\tt I}}, N_{\text{\tt O}}$ & input and output channel parallelism \\ 
        $k$ & number of MACs inside each S-MVE, $k \leq K_{\text{\tt x}}K_{\text{\tt y}}$ \\ 
    \bottomrule
    \end{tabular}
    \vspace{-0.3cm}
\end{table}

Elaborating on sparsity, it is the measure of the number of zero values in a stream of observed data. In this paper, we use $s_m$ to denote the instantaneous sparsity within the $m_{th}$ stream. 
The average sparsity is the expected value of the sparsity distribution ($\bar{s}_m = \mathbb{E} \left[ s_m \right]$). All these statistics are measured on a subset of ImageNet validation data.

In this section, the DSE problem of finding a maximal throughput hardware design for a given CNN model and FPGA pair is discussed.
Our DSE method decides the allocation of MACs based on the average sparsity (section \ref{subsec:mac}), as well as the insertion of buffers based on the variation of sparsity (section \ref{sec:buffer-sizing}).
Table~\ref{tab:symbols} gives the taxonomy of symbols used.

\subsection{MAC Allocation}
\label{subsec:mac}
In our design, all the MAC units are implemented with the DSP resources on the FPGA. As each convolutional layer is mapped to $N_{\text{\tt I}} \cdot N_{\text{\tt O}}$ S-MVE, each containing
$k$ MAC units, the per-layer DSP utilisation is modelled using the following equation.
\begin{equation}
    \mathcal{R}_{DSP} = N_{\text{\tt I}} \cdot N_{\text{\tt O}} \cdot k
\end{equation}

In streaming architectures, the average throughput of the whole system is dictated by the slowest layer. 
Therefore, the allocation of DSP resources is guided by the performance modelling of S-MVE. 
As each S-MVE is responsible for computing the kernel dimensions of the convolution, the average number of non-zero MAC operations required to produce one output is $(1-\bar{s}_{m}) \cdot K_{\text{\tt x}} \cdot K_{\text{\tt y}}$. 

As each S-VPE contains $k$ MACs, its average throughput can be expressed as, 

\begin{equation}
    \bar{\theta}_{m,n} = \min \left(1, \frac{k}{(1-\bar{s}_{m}) \cdot  K_{\text{\tt x}} \cdot  K_{\text{\tt y}}}\right)
    \label{equ:module_throughput}
\end{equation}

The performance increases as sparsity increases up to the maximum throughput of 1 element per cycle for each S-VPE. 
This model drives the sparsity-resource trade-off, where achieving maximum throughput does not necessarily require $K_{\text{\tt x}} \cdot K_{\text{\tt y}}$ MACs, and the saved resources can be allocated to other slower stages of the pipeline.
The S-MVE performance models can be used to construct the average latency, $\bar{t}_i$, of a convolutional layer,
\begin{equation}
    \bar{t}_i = H_{\text{\tt O}} \cdot W_{\text{\tt O}} \cdot \frac{C_{\text{\tt I}}}{N_{\text{\tt I}}} \cdot \frac{C_{\text{\tt O}}}{N_{\text{\tt O}}} \cdot \left( \max_{m \in [1, N_{\text{\tt I}}], n \in [1, N_{\text{\tt O}}]} \frac{1}{\bar{\theta}_{m,n}} \right)
    \label{equ:conv_lat}
\end{equation}

Given adequate buffering (as discussed in Section~\ref{sec:buffer-sizing}), the latency of the layer is dictated by the slowest S-MVE. Finally, the MACs allocation is expressed as the following optimisation problem:
\begin{equation}
    \max{\min_{i \in [1, L]} \frac{B}{\bar{t}_i}},\;\;s.t. \sum_{i}{\mathcal{R}_{DSP}} \leq budget
\end{equation}
where $L$ is the number of layers and $B$ is the batch size. This optimisation problem is solved using the simulated annealing algorithm \cite{montgomerie2022samo}.

\subsection{Buffer Depth Sizing}
\label{sec:buffer-sizing}
The above performance modelling in \eqref{equ:module_throughput} and \eqref{equ:conv_lat} assumes zero variance in each stream, which underestimates the latency given \textit{Jensen's inequality} that states $t(\mathbb{E}[\theta]) \leq \mathbb{E}[t(\theta)]$.
From the hardware perspective, latency underestimation is caused by back-pressure from the synchronisation barriers illustrated in Fig.~\ref{fig:streaming-diagram}, where the observed instantaneous sparsity deviates from its average value. 
It is therefore necessary to introduce buffering that reduces the \textit{Jensen gap} between estimated and actual latency.

Buffers are placed at the input of S-MVEs, accounting for variations in instantaneous sparsity between the S-MVE input streams.
In order to determine a suitable choice in buffer depth, a statistical method based on the calculation of a moving average of sparsity is proposed, which is given as,
\begin{equation}
    \psi_{m}^{w}(j) = \frac{1}{w} \sum_{i=j}^{j+w} s_m(i)
\end{equation}
$\psi_{m}^{w}$ is the time series for the moving average of stream $m$ for a window size of $w$, and $s_m$ is the time series of the sparsity observed on stream $m$.
The intuition is that as the buffer size increases, the detrended average value of samples in the buffers converge to the average sparsity level of the stream. 

To measure the impact of buffer sizing, the \textit{back pressure metric}, $\rho_w$ is proposed, which is defined as,
\begin{equation}
    \rho_w = \mathbb{E} \left[ \max_m \psi_{m}^{w} - \min_m \psi_{m}^{w} \right] - \left( \max_m \bar{s}_{m} - \min_m \bar{s}_m \right)
\end{equation}
This metric gives the average maximum difference between the moving average windows, reflecting the average number of extra samples needed to balance the workload in that period.
The difference in average sparsity between the most and least sparse streams is subtracted to normalise for unbalanced streams.
The greater $\rho_w$ is, the slower the execution of the hardware will be with respect to \eqref{equ:conv_lat}. 

\begin{figure}[t]
    \centering
    \resizebox{0.95\columnwidth}{!}{\definecolor{sparsecolor}{HTML}{b2df8a}
\definecolor{densecolor}{HTML}{1f78b4}
\definecolor{pgfjcred}{HTML}{e31a1c}

\begin{tikzpicture}
\pgfplotsset{
  width=0.8\columnwidth,
  height=50mm, 
  compat=1.5,
}
\begin{axis}[
    axis y line*=left,
    xlabel=Buffer Size ($w$),
    ylabel={LUTRAM Usage},
    xmin=0, xmax=256,
    ymin=0,
    xmajorgrids=true,
    minor tick num=1,
    xminorgrids=true,
    xtick={0,64,128,192,256},
    ytick={0, 1000, 2000},
    yticklabels={0, 1K, 2K},
    tick label style={font=\footnotesize},
]

\addplot [const plot, no marks, black, thick] 
coordinates {
    (0, 288)
    (32, 576)
    (64, 864)
    (96, 1152)
    (128, 1440)
    (160, 1728)
    (192, 2016)
    (224, 2304)
    (256, 2304)
};

\end{axis}
\begin{axis}[
    jcred,
    axis y line*=right,
    axis x line=none,
    ylabel={\textcolor{jcred}{Latency Overhead (\%)}},
    xmin=0, xmax=256,
    ymin=0, ymax=20,
    tick label style={font=\footnotesize},
]
\addplot [jcred, thick] 
coordinates {
    (2, 18.5524344569288)
    (4, 12.4606741573034)
    (6, 10.2883895131086)
    (8, 9.3876404494382)
    (10, 8.77153558052435)
    (12, 8.24906367041199)
    (14, 7.84082397003745)
    (16, 7.48314606741573)
    (18, 7.18164794007491)
    (20, 6.89138576779026)
    (22, 6.67790262172285)
    (24, 6.46629213483146)
    (26, 6.29213483146068)
    (28, 6.12734082397004)
    (30, 5.96067415730337)
    (32, 5.812734082397)
    (34, 5.65168539325843)
    (36, 5.52247191011236)
    (38, 5.4063670411985)
    (40, 5.30898876404494)
    (42, 5.21910112359551)
    (44, 5.12734082397004)
    (46, 5.03558052434457)
    (48, 4.94569288389513)
    (50, 4.8501872659176)
    (52, 4.76591760299625)
    (54, 4.67415730337079)
    (56, 4.5936329588015)
    (58, 4.52621722846442)
    (60, 4.45880149812734)
    (62, 4.4063670411985)
    (64, 4.35205992509363)
    (66, 4.29213483146067)
    (68, 4.24719101123596)
    (70, 4.20037453183521)
    (72, 4.16666666666667)
    (74, 4.10674157303371)
    (76, 4.06367041198502)
    (78, 4.00374531835206)
    (80, 3.95880149812734)
    (82, 3.90074906367041)
    (84, 3.84456928838951)
    (86, 3.80337078651685)
    (88, 3.76029962546817)
    (90, 3.71535580524345)
    (92, 3.65917602996255)
    (94, 3.61048689138577)
    (96, 3.55992509363296)
    (98, 3.52059925093633)
    (100, 3.49625468164794)
    (102, 3.46816479400749)
    (104, 3.44756554307116)
    (106, 3.41011235955056)
    (108, 3.37453183520599)
    (110, 3.3314606741573)
    (112, 3.30149812734082)
    (114, 3.25842696629214)
    (116, 3.22284644194757)
    (118, 3.18539325842697)
    (120, 3.15355805243446)
    (122, 3.11985018726592)
    (124, 3.08426966292135)
    (126, 3.04868913857678)
    (128, 3.00936329588015)
    (130, 2.97003745318352)
    (132, 2.92696629213483)
    (134, 2.88389513108614)
    (136, 2.83895131086142)
    (138, 2.79213483146067)
    (140, 2.74531835205993)
    (142, 2.69662921348315)
    (144, 2.64794007490637)
    (146, 2.59737827715356)
    (148, 2.55243445692884)
    (150, 2.52059925093633)
    (152, 2.48501872659176)
    (154, 2.45692883895131)
    (156, 2.41947565543071)
    (158, 2.39325842696629)
    (160, 2.36142322097378)
    (162, 2.32958801498127)
    (164, 2.30524344569288)
    (166, 2.26591760299625)
    (168, 2.24157303370787)
    (170, 2.20224719101124)
    (172, 2.17602996254682)
    (174, 2.15168539325843)
    (176, 2.12359550561798)
    (178, 2.0936329588015)
    (180, 2.0749063670412)
    (182, 2.04681647940075)
    (184, 2.04307116104869)
    (186, 2.04307116104869)
    (188, 2.04307116104869)
    (190, 2.04307116104869)
    (192, 2.04307116104869)
    (194, 2.04307116104869)
    (196, 2.04307116104869)
    (198, 2.04307116104869)
    (200, 2.04307116104869)
    (202, 2.04307116104869)
    (204, 2.04307116104869)
    (206, 2.04307116104869)
    (208, 2.04307116104869)
    (210, 2.04307116104869)
    (212, 2.04307116104869)
    (214, 2.04307116104869)
    (216, 2.04307116104869)
    (218, 2.04307116104869)
    (220, 2.04307116104869)
    (222, 2.04307116104869)
    (224, 2.04307116104869)
    (226, 2.04307116104869)
    (228, 2.04307116104869)
    (230, 2.04307116104869)
    (232, 2.04307116104869)
    (234, 2.04307116104869)
    (236, 2.04307116104869)
    (238, 2.04307116104869)
    (240, 2.04307116104869)
    (242, 2.04307116104869)
    (244, 2.04307116104869)
    (246, 2.04307116104869)
    (248, 2.04307116104869)
    (250, 2.04307116104869)
    (252, 2.04307116104869)
    (254, 2.04307116104869)
    (256, 2.04307116104869)
};
\end{axis}

\begin{axis}[
    jcblue, 
    axis y line*=right, 
    axis x line=none, 
    ylabel={\textcolor{jcblue}{Back Pressure Metric ($\rho_w$)}},
    xmin=0, xmax=256, 
    ymin=0, ymax=4,
    tick label style={font=\footnotesize},
]%
\pgfplotsset{
    every outer y axis line/.style={xshift=15mm}, 
    every tick/.style={xshift=15mm}, 
    every y tick label/.style={xshift=15mm},
}
\addplot [jcblue, thick] 
coordinates {
    (2, 3.7273829493379513)
    (4, 1.98294560978251)
    (6, 1.97423993307054)
    (8, 1.71603655857233)
    (10, 1.67980686094242)
    (12, 1.54519729266776)
    (14, 1.51781130627675)
    (16, 1.43785637346865)
    (18, 1.42960399344106)
    (20, 1.38502773152674)
    (22, 1.37072775946155)
    (24, 1.33354468531981)
    (26, 1.31314192213941)
    (28, 1.27846310777546)
    (30, 1.25594345501384)
    (32, 1.22375684008019)
    (34, 1.20318162896303)
    (36, 1.1747520902014)
    (38, 1.15415041885083)
    (40, 1.12754170398167)
    (42, 1.10662542392886)
    (44, 1.08125662504977)
    (46, 1.06059296994203)
    (48, 1.03645230935356)
    (50, 1.01642675820328)
    (52, 0.993601974896366)
    (54, 0.974218656111539)
    (56, 0.952578479521161)
    (58, 0.933900479537071)
    (60, 0.913202163599721)
    (62, 0.895012454965415)
    (64, 0.875010507042132)
    (66, 0.857383423438637)
    (68, 0.838154090155538)
    (70, 0.82122270944951)
    (72, 0.802894110263429)
    (74, 0.786871520105077)
    (76, 0.769542888692097)
    (78, 0.754359084677687)
    (80, 0.737966895317091)
    (82, 0.723871736840549)
    (84, 0.708561696762606)
    (86, 0.695242514723328)
    (88, 0.680880900869492)
    (90, 0.668282538652755)
    (92, 0.654674271300463)
    (94, 0.642822334921759)
    (96, 0.629816084574041)
    (98, 0.618497986308888)
    (100, 0.606240982004493)
    (102, 0.595616217158124)
    (104, 0.584325818589803)
    (106, 0.574481987090262)
    (108, 0.563853256305581)
    (110, 0.554605665252448)
    (112, 0.544664971851163)
    (114, 0.536027464402463)
    (116, 0.526828391819015)
    (118, 0.518927614512787)
    (120, 0.510455186475999)
    (122, 0.503104825267811)
    (124, 0.495329979173113)
    (126, 0.488459984731673)
    (128, 0.480926984000292)
    (130, 0.474414866362062)
    (132, 0.467394554387031)
    (134, 0.461281068945433)
    (136, 0.454630986465623)
    (138, 0.448936195450167)
    (140, 0.442798370935435)
    (142, 0.437499268269022)
    (144, 0.4316804404089)
    (146, 0.426609713977757)
    (148, 0.420982757720917)
    (150, 0.416169345032285)
    (152, 0.410801504853078)
    (154, 0.406298154718982)
    (156, 0.401221092087638)
    (158, 0.396881406875139)
    (160, 0.391993664416715)
    (162, 0.387782208157062)
    (164, 0.383021862939815)
    (166, 0.378980383521015)
    (168, 0.374398818644166)
    (170, 0.370475723243842)
    (172, 0.366002388101507)
    (174, 0.362114756888438)
    (176, 0.357718192665629)
    (178, 0.353912093300903)
    (180, 0.349656210113443)
    (182, 0.34605397160334)
    (184, 0.341957013801368)
    (186, 0.338419834796745)
    (188, 0.334485540171177)
    (190, 0.331046973368518)
    (192, 0.327290705275674)
    (194, 0.323961737989526)
    (196, 0.32032703140674)
    (198, 0.317120093087713)
    (200, 0.313617851408486)
    (202, 0.31059496208272)
    (204, 0.307247871674194)
    (206, 0.30433214979583)
    (208, 0.301169147667531)
    (210, 0.298397359171678)
    (212, 0.29531360415759)
    (214, 0.292696869787964)
    (216, 0.289810402556022)
    (218, 0.287383018675055)
    (220, 0.284716057622108)
    (222, 0.282424841130533)
    (224, 0.279825163949801)
    (226, 0.277627968312115)
    (228, 0.275163711473712)
    (230, 0.273154494470694)
    (232, 0.270846961367176)
    (234, 0.268977918292292)
    (236, 0.266827957790507)
    (238, 0.264998056715373)
    (240, 0.262954142013367)
    (242, 0.261223346407162)
    (244, 0.259263029596617)
    (246, 0.257789278529651)
    (248, 0.256029799980317)
    (250, 0.2547454292748)
    (252, 0.253193915186739)
    (254, 0.252032952325978)
    (256, 0.250637223163738)
};
\end{axis}

\end{tikzpicture}}
    \caption{
        Comparison of the back-pressure metric and observed latency overhead for different buffer sizes, for the $2^{\text{nd}}$ layer of ResNet-18 with a configuration of $N_I = 32$ and $k=1$. 
        The cost of the buffer in terms of LUTRAM is given for each buffer size.
    }
    \label{fig:buffer_size}
\end{figure}
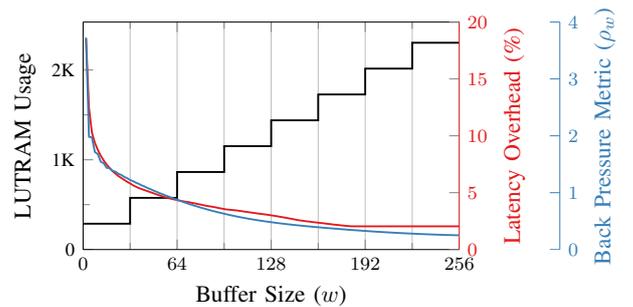

The effectiveness of the back pressure metric at identifying the optimal buffer size is evaluated in Fig.~\ref{fig:buffer_size}.
It can be seen that the metric is strongly correlated with the latency overhead observed.
The buffer size is chosen based on a stopping condition for $\rho_w$ as well as a limit on LUTRAM.

\begin{table*}[!t]
\setlength{\tabcolsep}{4pt}
\centering
\caption{
    Comparison of the performance and resources of related works for CNN models running ImageNet. 
    We use GOP/s/DSP to evaluate the performance with a normalised hardware resource since other approaches 
use different FPGA devices. 
    The \textcolor{forestgreen(web)}{\textbf{best}} result in each comparison is highlighted in bold and green.
    W. = weights, and P.-A. = post activation.
}
\label{tab:main-comparison}
\begin{tabular}{@{}c|cccccc|ccc|cc@{}}
\toprule
& \textbf{\cite{venieris_fpgaconvnet_2018}} & \textbf{\cite{li2021fpga}} & \textbf{\cite{lu2019efficient}} &  \textbf{\cite{zhu_efficient_2020}} & \textbf{Ours} & \textbf{Ours} & \textbf{\cite{venieris_fpgaconvnet_2018}} & \textbf{\cite{venieris2021unzipfpga}} & \textbf{Ours} & \textbf{\cite{lu2019efficient}} & \textbf{Ours}\\
\midrule
Network      & \multicolumn{6}{c|}{VGG16}               & \multicolumn{3}{c|}{ResNet-18} & \multicolumn{2}{c}{ResNet-50}\\  
\midrule
Quantisation  & W16A16 & W8A16 & W16A16 & W16A16 & W16A16 & W16A16   & W16A16  & W16A16   & W16A16 & W16A16 & W16A16\\
Sparsity      & -     & -    & W.    & W., P.-A.*    & P.-A.   & P.-A.         & -  & - & P.-A. & W.    & P.-A. \\
Streaming     & Yes    & No    & No     & No     & Yes   & Yes           & Yes & No & Yes  & No   & Yes\\
Device        & ZC706  & VC709 & ZCU102 & ZCU102 & ZC706 & ZCU102         & ZC706 & ZC706 & ZC706 & ZCU102  & ZCU102\\
Freq. (MHz)   & 200    & 200   & 200    & 200    & 200   & 200          & 200  & 150 & 200   & 200  & 200\\
LUT (k)       & 148 {\scriptsize (68\%)} & 121 {\scriptsize (28\%)} & 252 {\scriptsize (92\%)} & 178 {\scriptsize (65\%)} 
                &  \textcolor{forestgreen(web)}{\textbf{120} {\scriptsize (55\%)}} & 163 {\scriptsize (59\%)}
                & 147 {\scriptsize (67\%)} & 164 {\scriptsize (75\%)} 
                & \textcolor{forestgreen(web)}{\textbf{129} {\scriptsize (59\%)}} 
                & \textcolor{forestgreen(web)}{\textbf{252} {\scriptsize (92\%)}}  & 260 {\scriptsize (95\%)} \\
BRAM          & 798 {\scriptsize (73\%)} & 934 {\scriptsize (32\%)} & 912 {\scriptsize (50\%)} & 1460 {\scriptsize (80\%)} 
                & \textcolor{forestgreen(web)}{\textbf{504} {\scriptsize (46\%)}}  & 912 {\scriptsize (50\%)}   
                & \textcolor{forestgreen(web)}{\textbf{528} {\scriptsize (48\%)}}  &  948 {\scriptsize (87\%)} 
                & 586 {\scriptsize (54\%)} & \textcolor{forestgreen(web)}{\textbf{912} {\scriptsize (50\%)}}   
                & 1382 {\scriptsize (76\%)} \\
DSP           & 603 {\scriptsize (67\%)}  & 664 {\scriptsize (18\%)}  & 1144 {\scriptsize (45\%)}  
                & 1350 {\scriptsize (53\%)} & \textcolor{forestgreen(web)}{\textbf{512} {\scriptsize (57\%)}}  
                & 1024 {\scriptsize (41\%)} & 588 {\scriptsize (65\%)} & 900 {\scriptsize (100\%)} 
                & \textcolor{forestgreen(web)}{\textbf{528} {\scriptsize (59\%)}} 
                & 1144 {\scriptsize (45\%)} & \textcolor{forestgreen(web)}{\textbf{1032} {\scriptsize (41\%)}}\\
\midrule
GOP/s       & 198.0  & 230.1 & 309    & 495.4  & 310.8 & \textcolor{forestgreen(web)}{\textbf{534.4}}       & 135.0   & 181.6 & \textcolor{forestgreen(web)}{\textbf{185.4}}  & \textcolor{forestgreen(web)}{\textbf{291.4}} & 252.7 \\
GOP/s/DSP   & 0.33 & 0.35  & 0.27   & 0.37   & \textcolor{forestgreen(web)}{\textbf{0.61}} & 0.52 & 0.23 & 0.20 & \textcolor{forestgreen(web)}{\textbf{0.35}} & \textcolor{forestgreen(web)}{\textbf{0.25}}  & 0.24 \\
\bottomrule
\end{tabular}
\end{table*}

\section{Evaluation}
\label{sec:evaluation}

In this section, the performance and resource utilisation of the proposed framework is evaluated.
For hardware synthesis, Vivado 2020.1 is used. 

\subsection{Dense vs Sparse Design Comparison}
\label{subsec:dense_sparse_compare}

The performance benefits of sparsity on the proposed architecture are evaluated in this subsection.
The proposed toolflow is used to generate both dense designs using an existing Matrix-Vector Engine \cite{venieris_fpgaconvnet_2018}, as well as designs with the proposed Sparse Matrix-Vector Engine.

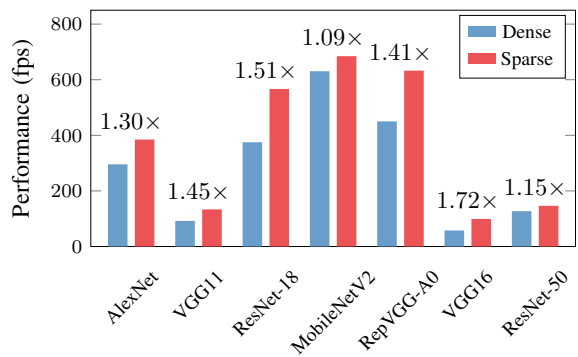
\begin{figure}[t]
    \centering
    \resizebox{0.92\columnwidth}{!}{\usetikzlibrary{calc}

\begin{tikzpicture}
\pgfplotsset{set layers, compat=1.5}
\begin{axis}[
    ybar=3pt,
    height=50mm,
    width=0.95\columnwidth,
    bar width=8,
    ymin=0, ymax=850,
    ticklabel style = {font=\small},
    ylabel={Performance (fps)},
    ylabel style = {align=center},
    xtick=data,
    xtick style={draw=none},
    xticklabels={AlexNet, VGG11, ResNet-18,
        MobileNetV2, RepVGG-A0, VGG16, ResNet-50},
    yticklabel style={font=\footnotesize},
    xticklabel style={rotate=45, font=\footnotesize},
    legend image code/.code={%
        \draw[#1] (0cm,-0.05cm) rectangle (0.4cm,0.1cm);
    },
  ]

\addplot [fill=jcblue!75, draw opacity=0, ] 
coordinates {
(1, 295.66) %
(2, 92.11) %
(3, 375.02) %
(4, 630.46) %
(5, 450.11) %
(6, 57.86) %
(7, 127.16) %
};\label{plot:dense}
\addlegendentry{\footnotesize Dense}

\addplot [fill=jcred!75, draw opacity=0, ] 
coordinates {
	(1, 384.86) %
	(2, 133.49) %
	(3, 566.71) %
	(4, 684.62) %
	(5, 632.61) %
	(6, 99.05) %
	(7, 146.38) %
}; \label{plot:sparse}
\addlegendentry{\footnotesize Sparse}

\node[above] at ($(axis cs:1, 384.86)$) {$1.30\times$};
\node[above] at ($(axis cs:2, 134)$) {$1.45\times$};
\node[above] at ($(axis cs:3, 567)$) {$1.51\times$};
\node[above] at ($(axis cs:4, 685)$) {$1.09\times$};
\node[above] at ($(axis cs:5, 633)$) {$1.41\times$};
\node[above] at ($(axis cs:6, 99.05)$) {$1.72\times$};
\node[above] at ($(axis cs:7, 146.38)$) {$1.15\times$};

\end{axis}
\end{tikzpicture}}
    \caption{
        Performance comparison between the dense streaming accelerator and the proposed sparse streaming accelerators for representative CNN workloads, targeting a U250 FPGA. 
        Our tool generates designs with up to 50\% greater performance.
    }
    \label{fig:sparse-dense-comparison}
\end{figure}

Fig.~\ref{fig:sparse-dense-comparison} illustrates the dense and sparse hardware performance for a set of representative CNN workloads: AlexNet \cite{alexnet}, VGG11 \& VGG16 \cite{vgg}, RepVGG-AO \cite{ding2021repvgg}, MobilenetV2 \cite{howard2017mobilenets} and ResNet-18 \& ResNet-50 \cite{resnet}. 
The sparse engine exceeds the performance of the dense engine for all CNN models. 
The largest gain was observed for ResNet-18 ($51\%$), whereas the smallest was for MobileNetv2 ($9\%$).
For both MobileNetV2 and ResNet-50, the performance gains realised were marginal. 
In the case of MobileNetV2, this is due to most of the workload being point-wise convolutions, as the proposed S-MVE hardware is not able to exploit the sparsity of $1\times1$ kernels.
For ResNet-50, both dense and sparse designs are constrained by the large on-chip memory requirements, bottlenecking their achievable performance.
Overall, the proposed toolflow is able to exploit the post-activation sparsity leading to significant performance gains.

The performance improvements for the sparse architecture can be attributed to sparsity.
For example, the average sparsity across all the convolutional layers of VGG16 and ResNet-18 is 0.65 and 0.57 respectively for the ImageNet validation dataset.
This suggests $1\!/\!(1\!-\!0.65)\!=\!2.86$ and $1\!/\!(1\!-\!0.57)\!=\!2.32$ speed-up at maximum. 
The gap between the theoretical maximum speed-up and the improvement achieved is due to the overhead on clock frequency and LUT usage, as described in Section~\ref{sec:smve-hardware}. 

A detailed comparison between the dense and sparse engines focusing on the 3rd convolutional layer of VGG16 is given in Table~\ref{tab:dense-sparse-detailed}, which is representative for many $3\times3$ convolutional layers across CNN models.
The results demonstrate the performance benefits of the proposed S-MVE module, as it is able to effectively bypass zero multiplications, reducing the latency by 60\%.
At the same time, the clock frequency and LUT usage are penalised by 10\% and 50\%  receptively, creating different bottlenecks during the DSE for the entire network.
\begin{table}[h]
    \centering
    \caption{
        A case study comparing dense and sparse architectures for the $3^{\text{rd}}$ convolutional layer of VGG16. 
        The sparse hardware achieves better performance with the overhead on LUT, FF and frequency.
    }
    \label{tab:dense-sparse-detailed}
    \begin{tabular}{ccccccc}
    \toprule
    \textbf{Design} & \textbf{LUT} & \textbf{FF} & \textbf{BRAM} & \textbf{DSP} & \begin{tabular}[c]{@{}c@{}}\textbf{Freq.}\\ (MHz)\end{tabular} & \begin{tabular}[c]{@{}c@{}}\textbf{Lat.}\\ (ms)\end{tabular} \\
    \midrule
        \textit{Dense}  & 26,046         & 41,211        & 272           & 192           & 223           & 44.5 \\
        \textit{Sparse} & 38,112         & 48,895        & 272           & 192           & 200           & 17.8 \\
               & \textcolor{jcred}{\textbf{1.5$\times$}}  & \textcolor{jcred}{\textbf{1.2$\times$}}     & 
               \textbf{\textcolor{jcgreen}{1.0$\times$}}   & \textbf{\textcolor{jcgreen}{1.0$\times$}}    & 
               \textcolor{jcred}{\textbf{0.9$\times$}}    & \textcolor{jcgreen}{\textbf{0.4$\times$}} \\
    \bottomrule
    \end{tabular}
\end{table}

\subsection{Comparing with Existing Sparse Accelerators}

In Table~\ref{tab:main-comparison}, our work is evaluated against both instruction-based sparse accelerators \cite{lu2019efficient, zhu_efficient_2020}, as well as a state-of-the-art streaming but dense accelerator \cite{venieris_fpgaconvnet_2018}. 
For sparse works, we achieved up to 1.93$\times$ GOP/s/DSP on VGG16, which demonstrates the benefit of combining post-activation sparsity exploitation and streaming. 
Compared to a high-performance streaming architecture \cite{venieris_fpgaconvnet_2018}, our work achieves 1.85$\times$ and 1.52$\times$ GOP/s/DSP on VGG16 and ResNet-18 respectively, without any degradation on the network accuracy. 
This reiterates the impact sparsity has on performance.

Our hardware is not able to outperform \cite{lu2019efficient} on ResNet-50 due to the limited LUT resources. 
We observe that both our design and the design in \cite{lu2019efficient} are LUT-bounded, however, compared with their instruction-based architecture, streaming architectures require extra buffers for weight storage and pipelining between layers, consuming additional LUTRAM and BRAM.
From table~\ref{tab:main-comparison}, our design uses 4\% less DSP than \cite{lu2019efficient}, yet consumes 3\% more LUT and 26\% more BRAM comparatively. 
Therefore, in order to fully exploit the potential of our design, devices with more on-chip memory resources are desirable.

\section{Conclusion}
\label{sec:conclusion}

In this work, a toolflow is proposed for exploiting post-activation sparsity in streaming-based CNN accelerators.
We address the key challenges which arise from non-deterministic sparse execution including dynamic scheduling, data stream synchronisation and statistics-aware design space exploration.
Overall, our method can achieve 1.41$\times$ to 1.93$\times$ greater performance compared to existing instruction-based  sparse accelerators.
With regard to future work, we are exploring the opportunity of CNN-accelerator co-design, such as encouraging input-sparsity of the slowest layer in the pipeline through a sparsity regulariser.

\section*{Acknowledgements}
For the purpose of open access, the authors have applied a Creative Commons Attribution (CC BY) license to any Accepted Manuscript version arising.

\bibliographystyle{IEEEtran}
\bibliography{bibliography}

\end{document}